\documentclass[twocolumn,APL,superscriptaddress]{revtex4}
\usepackage{bm}
\usepackage{amssymb}
\usepackage{amsmath}
\usepackage{graphicx}
\usepackage[sort&compress]{natbib}
\usepackage{epsfig}
\usepackage{color}

\usepackage[colorlinks=true, pagebackref=false, bookmarks=true,bookmarksopen=true,bookmarksnumbered=true]{hyperref}
\newcommand{\Cu}{\ensuremath{\mathrm{Cu}}}

\newcommand{\Nb}{\ensuremath{\mathrm{Nb}}}
\newcommand{\Al}{\ensuremath{\mathrm{Al}}}
\newcommand{\Si}{\ensuremath{\mathrm{Si}}}
\renewcommand{\O}{\ensuremath{\mathrm{O}}}

\renewcommand{\Re}{\ensuremath{\mathrm{Re}}}
\newcommand{\N}{\ensuremath{\mathrm{N}}}
\newcommand{\Ti}{\ensuremath{\mathrm{Ti}}}

\renewcommand{\section}[1]{}
\renewcommand{\subsection}[1]{}
\newcommand{\W}[1]{{#1}}

\begin{document}

\title{Coherence in a transmon qubit with epitaxial tunnel junctions \footnote{This contribution of NIST, an agency of the U.S. Government, is not subject to copyright.}%
}

\author{Martin P. Weides}
\email{martin.weides@NIST.gov}
\affiliation{National Institute of Standards and Technology, Boulder, Colorado 80305, USA}

\author{Jeffrey S. Kline}
\affiliation{National Institute of Standards and Technology, Boulder, Colorado 80305, USA}

\author{Michael R. Vissers}
\affiliation{National Institute of Standards and Technology, Boulder, Colorado 80305, USA}

\author{Martin O. Sandberg}
\affiliation{National Institute of Standards and Technology, Boulder, Colorado 80305, USA}

\author{David S. Wisbey}
\altaffiliation{Present address: Saint Louis University, St. Louis, Missouri 63103, USA}
\affiliation{National Institute of Standards and Technology, Boulder, Colorado 80305, USA}

\author{Blake R. Johnson}
\affiliation{Raytheon BBN Technologies, Cambridge, Massachusetts 02138, USA}

\author{Thomas A. Ohki}
\affiliation{Raytheon BBN Technologies, Cambridge, Massachusetts 02138, USA}

\author{David P. Pappas}
\email{david.pappas@NIST.gov}
\affiliation{National Institute of Standards and Technology, Boulder, Colorado 80305, USA}

\date{\today}

\begin{abstract}
We developed transmon qubits based on epitaxial tunnel junctions and interdigitated capacitors. This multileveled qubit, patterned by use of all-optical lithography, is a step towards scalable qubits with a high integration density. The relaxation time $T_1$ is $.72-.86\;\rm{\mu sec}$ and the ensemble dephasing time $T_2^*$ is slightly larger than $T_1$. The dephasing time $T_2$ ($1.36\;\rm{\mu sec}$) is nearly energy-relaxation-limited. Qubit spectroscopy yields weaker level splitting than observed in qubits with amorphous barriers in equivalent-size junctions. The qubit's inferred microwave loss closely matches the weighted losses of the individual elements (junction, wiring dielectric, and interdigitated capacitor), determined by independent resonator measurements.
\end{abstract}
\pacs{%
  74.50.+r, 
 85.25.Cp, 
 85.25.-j 	
}

\keywords{%
  Josephson junctions, superconducting qubit, dielectric loss
}

\maketitle

Quantum information processing receives considerable interest due to the potential speedup over classical information processing. Among the proposals and architectures, solid-state devices have the advantage of large-scale integration and flexibility in layout. In recent years, great experimental progress has been achieved by use of superconducting qubits. Operations such as control, coupling, and readout have made remarkable experiments possible, e.g., violation of Bell's inequality \cite{AnsmannNature09,Chow_PRA10_Bell}, three-qubit entanglement \cite{NeeleyNature10,DiCarlo_Nat10}, and quantum non-demolition readout \cite{LupascuQND07,Johnson_NaturePhys10}. Among the variety of superconducting qubits being proposed and realized, the transmon \cite{Koch_TransmonPRA07} provides dispersive readout, tunability, and first-order insensitivity against flux noise at the sweet spot, with a simple layout.

So far, transmons have been realized as single layer devices \cite{SchreierPRB08,Vijay_PRL11,Houck_PRL08} using Dolan-bridge tunnel junctions \cite{Dolan77} with energy relaxation times, $T_1$, ranging from a few hundred nanoseconds \cite{Vijay_PRL11} to a few microseconds \cite{Houck_PRL08}. However, their fast and simple fabrication process has difficulties in scalability and high integration density compared to multileveled (e.g . multiple patterning layers) qubits due to the absence of wiring crossovers and vias. The standard multileveled qubits, based on amorphous $\Al\O_x$ tunnel barriers with $\Al$ or $\Nb$ electrodes and micrometer-sized junctions, suffer from detrimental interaction with two-level-system (TLS) defects inside the amorphous barrier or crossover wiring dielectric that can absorb energy and interfere with the qubit state  \cite{SimmondsPRL04,MartinisPRL05}, resulting in shorter coherence times. By using epitaxial tunnel barriers, which have fewer defect states than amorphous barriers \cite{KlineSST09,Oh_PRB06}, we aim to improve scalability while maintaining the qubit coherence.

In this paper we present a multileveled transmon based on epitaxial materials and all-optical litho\-graphy. While the qubit spectroscopy shows weak avoided level crossings (maximum coupling strength $ 7\;\rm{MHz}$), presumably due to coupling to TLS, both the relaxation time $T_1$ and dephasing time $T_2$ exceed the best reported values in other multileveled qubits \cite{KlineSST09,Weides_Trilayer,MartinisPRL05}. The qubit's loss agrees with a upper bound value estimated from the weighted individual component's microwave losses, independently determined by separate measurements on notch-type resonators with $\Al_2\O_3$ parallel plate capacitors or coplanar waveguides (CWP).

\begin{figure}[tb]
\begin{center}
\includegraphics[width=8.6cm]{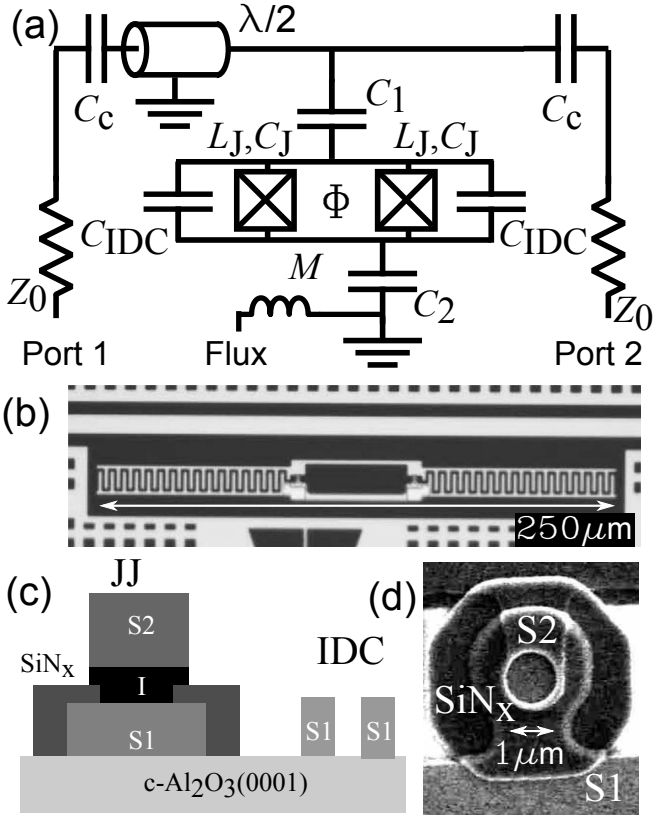}
\caption{(a) Circuit diagram of transmon (split JJ shunted with IDCs) coupled to $\lambda/2$ resonator. (b) Optical micrograph shows cavity resonator (top), transmon qubit (middle) and part of flux bias (bottom). All features are surrounded by flux holes in S1. (c) Cross-section (not to scale) of Josephson junction (JJ) and interdigitated capacitor (IDC). (d) SEM micrograph of $1.1\;\rm{\mu m^2}$ JJ with minimal $\Si\N_x$ insulation.}\label{Fig_1}
\end{center}
\end{figure}

Our qubit is depicted in Fig. \ref{Fig_1}. It is based on an epitaxial $\Re_{110\rm{nm}}/(\Ti_{1.5\rm{nm}}/\Re_{10\rm{nm}})_{2}\Ti_{1.5\rm{nm}}\Re_{15\rm{nm}}$ stack (labeled as $\Re/\Ti$) grown on a $76\;\rm{mm}$ diameter c-plane sapphire substrate. This $\Re/\Ti$ multilayer forms the bottom electrode (S1). Its smooth surface, $\sim1\;\rm{nm}$ rms roughness, compared to a bare $\Re$ electrode, $\geq 3\;\rm{nm}$, of similar thickness \cite{OhEpitaxialReSputtering}, is a basic requirement for the growth of uniform tunnel barriers. The transmon consists of a split Josephson junction (JJ) (drawn area per JJ: $1\;\rm{\mu m^2}$, electrical area determined via scanning electron microscopy: $1.1\;\rm{\mu m^2}$) with each junction having a $30\;\rm{fF}$ shunting interdigitated capacitor $C_{\rm{IDC}}$ (see Fig. \ref{Fig_1} b). The JJ area is $\sim25$ times larger than the conventional transmon \cite{SchreierPRB08}, and similar to the standard phase qubit junction areas \cite{Weides_Trilayer}. The flux-threaded loop is $15\times 50\;\rm{\mu m^2}$. The JJs are via-style junctions embedded in $\sim250\;\rm{nm}$ thick PECVD grown $\Si\N_x$ formed by a process that not only provides crystalline tunnel barriers, but defines small junction areas without perimeter defects \cite{Kline_SUST11}. We minimized the $\Si\N_x$ overlap region $C_{\rm{iso}}$ as much as possible to reduce the dielectric loss participation (see Fig. \ref{Fig_1} d). We grew the $(0001)$-oriented $\Al_2\O_3$ tunnel barrier (labeled I) at $900^{\circ} \rm{C}$. The counter-electrode (S2) is formed by room-temperature-deposited aluminum and is moderately textured and in-plane crystalline ordered: see cross-section view in Fig. \ref{Fig_1} c. Most structures are patterned in S1, except one half of the split JJ loop is formed by S2. The multileveled fabrication allows the use of overlaps and vias, important elements for scalable qubits with a high integration density.

The qubit state is read out dispersively via a half-wavelength resonator. The floating qubit island is capacitively coupled (effective coupling capacitor $C_{\rm{g}}=1/(C_{\rm{1}}^{-1}+C_{\rm{2}}^{-1})\sim7\;\rm{fF}$) to both the resonator ($C_{\rm{1}}$) and the ground ($C_2$), as shown in Fig. \ref{Fig_1} a. The meandered resonator ($f_r=8.3\;\rm{GHz}$) is coupled to the transmission lines via coupling capacitors $C_{\rm{c}}\sim5\;\rm{fF}$ (designed coupling $Q=10\;000$), setting the average photon loss rate to $\kappa/2\pi\sim0.8\;\rm{MHz}$. The designed vacuum Rabi frequency $g/2\pi$ is $85\;\rm{MHz}$. The qubit is placed at $\lambda/20$ from one resonator port to make inductive coupling negligible. The split junction is flux-biased via an on-chip inductor (mutual inductance $M=1.3\;\rm{pH}$) formed by a $50\;\rm{\Omega}$ coplanar waveguide terminated with an off-centered and 'T'-shaped inductive short to ground to minimize the cross-talk to the cavity. We estimate that coupling to the flux bias impedance limits the qubit lifetime to about $50\;\rm{\mu sec}$, well above the measured relaxation time. The Purcell limited relaxation time is calculated as $\left(\Delta/g\right)^2 /\kappa\approx 27\;\rm{\mu sec}$ at the sweet spot (detuning of $\Delta/2\pi\approx1\;\rm{GHz}$) \cite{Koch_TransmonPRA07,Houck_PRL08}.

The $\Re/\Ti\textrm{-}\Al_2\O_3\textrm{-}\Al$ epitaxial tunnel junctions have a room-temperature resistance $\times$ drawn area product $RA$ of $\sim2\textrm{-}4\;\rm{k\Omega \mu m^2}$. Sub-micrometer junctions have slightly larger values due to process bias. The barrier material loss tangent, $6\cdot 10^{-5}$, is estimated from multiplexed notch-type LC resonators with $6.0\;\rm{nm}$ thick $\Al_2\O_3$ parallel plate capacitors of  $100\;\rm{\mu m^2}$ area. These thick $\Al_2\O_3$ films show a lower degree of surface structure (measured by RHEED) than the $1.8\;\rm{nm}$ thin tunnel barrier, which may be reflected in the  degradation of its loss tangent. The junction's specific capacitance, $\sim60\;\rm{fF/\mu m^2}$, is inferred from the qubit anharmonicity and matches the dielectric constant of $\Al_2\O_3$. The $\Si\N_x$ loss tangent of $1\cdot10^{-3}$ was determined from the participation factor of thick films covering $\lambda/2$ CPW resonators. The $\Re/\Ti$ multilayer's microwave loss was measured in CPW resonators. As in the qubit, these S1 features were patterned first, and hence exposure to subsequent depositions and etches increased their internal single photon loss to $4\cdot10^{-5}$. Later, we noted that resonators instead patterned in the final step exhibit no such processing-induced increased loss.

\begin{figure}[tb]
\begin{center}
\includegraphics[width=8.6cm]{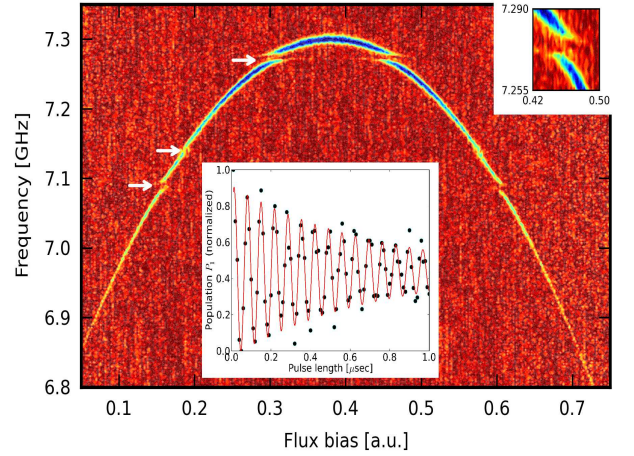}
\caption{The qubit spectroscopy shows some avoided level crossings (arrows). They are stable over the timescale of days, but change position after thermal cycling to room temperature. Maximum observed splitting size is $7\;\rm{MHz}$. Upper inset: Level splitting at $7.273\;\rm{GHz}$. Lower inset: Vacuum Rabi oscillations at sweet spot. The solid line represents a fit to the data.}\label{Fig_2}
\end{center}
\end{figure}

We measured the qubit in a dilution refrigerator at $25\;\rm{mK}$ using homodyne detection and a HEMT amplifier at $4\;\rm{K}$. The chip is thermally anchored to a $\Cu$ block covered with an $\Al$ lid, magnetically shielded with \W{a light-tight cryoperm cover coated with microwave absorbing material} and bolted onto the mixing stage \cite{Corcoles_APL_2011}. It is electrically connected via bond wires to a printed-circuit board and the microwave and bias lines.

From microwave spectroscopy, we determine the maximum qubit frequency $\omega_{10}/2\pi=7.3\;\rm{GHz}$, being $1\;\rm{GHz}$ detuned from the fundamental cavity mode. \W{We determine the charging energy $E_C/h=97\;\rm{MHz}$ by the device anharmonicity of $100\;\rm{MHz}$ via single-tone spectroscopy versus power} and $E_J/E_C=720$ ($RA=6.5\;\rm{k\Omega \mu m^2}$) at the sweet spot frequency. The total qubit capacitance is $C_{\sum}=200\;\rm{fF}$. Fig. \ref{Fig_2} shows the qubit spectroscopy, level splitting (upper inset) and vacuum Rabi oscillations (lower inset). \W{We note three avoided level crossings, which we attribute to the presence of TLSs. The maximum coupling strength, measured on two qubits in four cooldowns, was $7$ MHz. }The anticrossings show no flux dependence, as expected for electronic defect states. While they are stable over several days, they change freqency and splitting size under thermal cycling. The qubit parameters of one qubit did not change over three cooldowns.
Transmons are well suited for TLS spectroscopy, as the maximal TLS coupling strength to the qubit is $\propto 1/\sqrt{C_{\sum}}$ \cite{MartinisPRL05}.  \W{The observed TLSs are stronger dipole coupled to a transmon qubit than to a phase qubit with $\sim$ 6 times larger capacitance and a comparable JJ area \cite{Weides_Trilayer}. In phase qubits the TLS coupling strength would be on the order of the $\sim 2\;\rm{MHz}$ typical qubit linewidth} and would be unresolvable by standard spectroscopy \cite{AllmanArXiv}.

\begin{figure}[tb]
\begin{center}
\includegraphics[width=8.6cm]{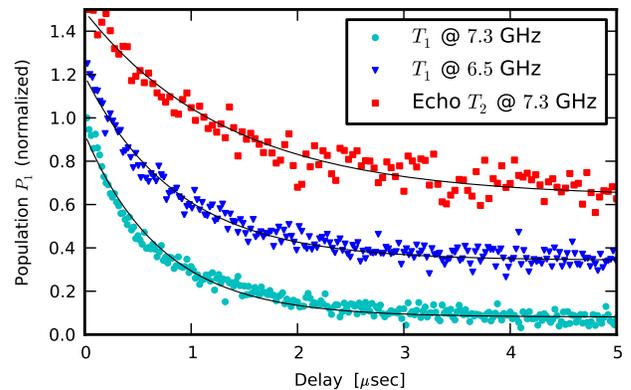}
\caption{Qubit population in time domain: $T_1=0.73\;\rm{\mu sec}$ and echo $T_2=1.36\;\rm{\mu sec}$ at the sweet spot, and $T_1=0.86\;\rm{\mu sec}$ (detuned). Successive time traces have been offset vertically for clarity. The lines represent fits to the data. }\label{Fig_3}
\end{center}
\end{figure}

Qubit time domain measurements (Fig. \ref{Fig_3}) at the flux sweet spot yield a relaxation time $T_1\approx 0.73 \;\rm{\mu sec}$, ensemble dephasing time $T_2^*\approx 0.92 \;\rm{\mu sec}$ (including low-frequency noise) and echo-corrected dephasing $T_2\approx 1.36 \;\rm{\mu sec}$.  Detuning by $800\;\rm{MHz}$ increases $T_1$ slightly to $\approx 0.86 \;\rm{\mu sec}$. We attribute this variation to stochastic effects, as the Purcell limit is considerably larger.  From the measured $T_1$ we determine the loss tangent of the qubit to be $\delta_m = (T_1\;\omega_{10})^{-1} \approx 3\cdot10^{-5}$ at a qubit frequency $\omega_{10}/2\pi=7.3\;\rm{GHz}$.

The calculated weighted loss tangent from table \ref{Tab:LossEstimation} of a parallel combination of capacitors is given by $\tan{\delta}=\sum_{i}C_i \tan\delta_i/C_{\sum}=5.7\cdot10^{-5}$. \W{This weighted loss sets an upper estimate on the effective qubit loss} \cite{MartinisPRL05} and matches the measured loss $\delta_m$ within a factor of 2, which is quite close considering the residual uncertainties in resonator loss and capacitance determination of the individual elements.

The ensemble dephasing time $T_2^*$ in our epitaxial qubit is slightly larger than the relaxation time $T_1$. The large $C_{\sum}$ renders the transmon insensitive to $1/f$-charge noise as the charge dispersion is suppressed, according to $\sim\exp{(-\sqrt{E_J/E_C})}$ \cite{Koch_TransmonPRA07,SchreierPRB08}. At the sweet spot, the qubit's dephasing is relatively weakly affected by other sources, such as critical-current noise (first-order effect) and external magnetic field fluctuations coupled to the relatively large loop or flux noise in the epitaxial materials (second-order effects). As $T_2^*$ is of the order of $T_1$, the qubit is nearly homogeneously broadened. The lower-frequency noise affects the qubit dephasing rate to the same extent as conventional $\Al$ transmons \cite{Houck_PRL08}.

\begin{table}[b]
\caption{\label{Tab:LossEstimation} Overview on single-photon loss tangent, capacitance and participation for the individual elements. The losses for $\Al_2\O_3$ tunnel barrier,  $\Si\N_x$ overlaps, and $\Re/\Ti$  IDCs are inferred from lumped LC and CPW resonators. Elements with a small $C_i/C_{\sum}\tan\delta_i$ contribution are neglected, e.g., the split JJ loop. The calculated effective weighted transmon loss tangent is $5.7\cdot10^{-5}$.}

\begin{tabular}{l  l  l  l  l}
  \hline
  \hline
  Capacitive element & &  measured $\delta_i$& $C_i$&$\frac{C_i}{C_{\sum}}\tan\delta_i$\\
  & &&$[\rm{fF}$] & \\
  \hline
  $\Al_2\O_3$  barrier &$C_{\rm{J}}$  & $6\cdot10^{-5}$& 132 &$4.0\cdot10^{-5}$\\
  $\Si\N_x$  insulation  &$C_{\rm{iso}}$ &  $1\cdot10^{-3}$ & 0.8& $3.8\cdot10^{-6}$\\
  $\Re/\Ti$ on substrate &$2C_{\rm{IDC}}+C_{\rm{g}}$ & $4\cdot10^{-5}$  & 67 &$1.3\cdot10^{-5}$\\
  \hline
  weighted &$C_{\sum}$ & $$  & 200 &$5.7\cdot10^{-5}$\\
  \hline
  \hline
\end{tabular}\end{table}

In conclusion, we have fabricated and characterized an epitaxial multileveled transmon qubit. Both $T_1$ and $T_2$ exceed the best reported values for multileveled qubits. The measured relaxation time matches well the weighted loss contributions of the individual elements. The residual TLSs coupling strength is reduced, compared to qubits with amorphous barriers in equivalent-size junctions. These results were qualitatively repeated after several months and verified on a second sample with slightly different design parameters.

The good agreement between the microwave loss of the qubit and the individual elements should allow us to systematically improve the relaxation time $T_1$ in our future devices by (i) reducing the IDC loss and (ii) implementing smaller junctions.\\
The authors gratefully acknowledge the assistance of F. Farhoodi and valuable discussions with F. da Silva and R. Simmonds. This work was funded by by the US government, Office of the Director of National Intelligence (ODNI), Intelligence Advanced Research Projects Activity (IARPA),  and NIST Quantum Information initiative. All statements of fact, opinion, or conclusions contained herein are those of the authors and should not be construed as representing the official views or policies of ODNI or IARPA.


\newpage

\begin{thebibliography}{19}
\expandafter\ifx\csname natexlab\endcsname\relax\def\natexlab#1{#1}\fi
\expandafter\ifx\csname bibnamefont\endcsname\relax
  \def\bibnamefont#1{#1}\fi
\expandafter\ifx\csname bibfnamefont\endcsname\relax
  \def\bibfnamefont#1{#1}\fi
\expandafter\ifx\csname citenamefont\endcsname\relax
  \def\citenamefont#1{#1}\fi
\expandafter\ifx\csname url\endcsname\relax
  \def\url#1{\texttt{#1}}\fi
\expandafter\ifx\csname urlprefix\endcsname\relax\def\urlprefix{URL }\fi
\providecommand{\bibinfo}[2]{#2}
\providecommand{\eprint}[2][]{\url{#2}}

\bibitem[{\citenamefont{Ansmann \emph{et~al.}}(2009)\citenamefont{Ansmann,
  Wang, Bialczak, Hofheinz, Lucero, Neeley, O'Connell, Sank, Weides, Wenner,
  Cleland, and Martinis}}]{AnsmannNature09}
\bibinfo{author}{\bibfnamefont{M.}~\bibnamefont{Ansmann}},
  \bibinfo{author}{\bibfnamefont{H.}~\bibnamefont{Wang}},
  \bibinfo{author}{\bibfnamefont{R.~C.} \bibnamefont{Bialczak}},
  \bibinfo{author}{\bibfnamefont{M.}~\bibnamefont{Hofheinz}},
  \bibinfo{author}{\bibfnamefont{E.}~\bibnamefont{Lucero}},
  \bibinfo{author}{\bibfnamefont{M.}~\bibnamefont{Neeley}},
  \bibinfo{author}{\bibfnamefont{A.~D.} \bibnamefont{O'Connell}},
  \bibinfo{author}{\bibfnamefont{D.}~\bibnamefont{Sank}},
  \bibinfo{author}{\bibfnamefont{M.}~\bibnamefont{Weides}},
  \bibinfo{author}{\bibfnamefont{J.}~\bibnamefont{Wenner}},
  \bibinfo{author}{\bibfnamefont{A.~N.} \bibnamefont{Cleland}},
  \bibnamefont{and} \bibinfo{author}{\bibfnamefont{J.~M.}
  \bibnamefont{Martinis}}, \bibinfo{journal}{Nature (London)}
  \textbf{\bibinfo{volume}{461}}, \bibinfo{pages}{504} (\bibinfo{year}{2009}).

\bibitem[{\citenamefont{Chow \emph{et~al.}}(2010)\citenamefont{Chow, DiCarlo,
  Gambetta, Nunnenkamp, Bishop, Frunzio, Devoret, Girvin, and
  Schoelkopf}}]{Chow_PRA10_Bell}
\bibinfo{author}{\bibfnamefont{J.~M.} \bibnamefont{Chow}},
  \bibinfo{author}{\bibfnamefont{L.}~\bibnamefont{DiCarlo}},
  \bibinfo{author}{\bibfnamefont{J.~M.} \bibnamefont{Gambetta}},
  \bibinfo{author}{\bibfnamefont{A.}~\bibnamefont{Nunnenkamp}},
  \bibinfo{author}{\bibfnamefont{L.~S.} \bibnamefont{Bishop}},
  \bibinfo{author}{\bibfnamefont{L.}~\bibnamefont{Frunzio}},
  \bibinfo{author}{\bibfnamefont{M.~H.} \bibnamefont{Devoret}},
  \bibinfo{author}{\bibfnamefont{S.~M.} \bibnamefont{Girvin}},
  \bibnamefont{and} \bibinfo{author}{\bibfnamefont{R.~J.}
  \bibnamefont{Schoelkopf}}, \bibinfo{journal}{Phys. Rev. A}
  \textbf{\bibinfo{volume}{81}}, \bibinfo{pages}{062325}
  (\bibinfo{year}{2010}).

\bibitem[{\citenamefont{Neeley \emph{et~al.}}(2010)\citenamefont{Neeley,
  Bialczak, Lenander, Lucero, Mariantoni, O'Connell, Sank, Wang, Weides,
  Wenner, Yin, Yamamoto, Cleland, and Martinis}}]{NeeleyNature10}
\bibinfo{author}{\bibfnamefont{M.}~\bibnamefont{Neeley}},
  \bibinfo{author}{\bibfnamefont{R.~C.} \bibnamefont{Bialczak}},
  \bibinfo{author}{\bibfnamefont{M.}~\bibnamefont{Lenander}},
  \bibinfo{author}{\bibfnamefont{E.}~\bibnamefont{Lucero}},
  \bibinfo{author}{\bibfnamefont{M.}~\bibnamefont{Mariantoni}},
  \bibinfo{author}{\bibfnamefont{A.~D.} \bibnamefont{O'Connell}},
  \bibinfo{author}{\bibfnamefont{D.}~\bibnamefont{Sank}},
  \bibinfo{author}{\bibfnamefont{H.}~\bibnamefont{Wang}},
  \bibinfo{author}{\bibfnamefont{M.}~\bibnamefont{Weides}},
  \bibinfo{author}{\bibfnamefont{J.}~\bibnamefont{Wenner}},
  \bibinfo{author}{\bibfnamefont{Y.}~\bibnamefont{Yin}},
  \bibinfo{author}{\bibfnamefont{T.}~\bibnamefont{Yamamoto}},
  \bibinfo{author}{\bibfnamefont{A.~N.} \bibnamefont{Cleland}},
  \bibnamefont{and} \bibinfo{author}{\bibfnamefont{J.~M.}
  \bibnamefont{Martinis}}, \bibinfo{journal}{Nature (London)}
  \textbf{\bibinfo{volume}{467}}, \bibinfo{pages}{570} (\bibinfo{year}{2010}).

\bibitem[{\citenamefont{DiCarlo \emph{et~al.}}(2010)\citenamefont{DiCarlo,
  Reed, Sun, Johnson, Chow, Gambetta, Frunzio, Girvin, Devoret, and
  Schoelkopf}}]{DiCarlo_Nat10}
\bibinfo{author}{\bibfnamefont{L.}~\bibnamefont{DiCarlo}},
  \bibinfo{author}{\bibfnamefont{M.~D.} \bibnamefont{Reed}},
  \bibinfo{author}{\bibfnamefont{L.}~\bibnamefont{Sun}},
  \bibinfo{author}{\bibfnamefont{B.~R.} \bibnamefont{Johnson}},
  \bibinfo{author}{\bibfnamefont{J.~M.} \bibnamefont{Chow}},
  \bibinfo{author}{\bibfnamefont{J.~M.} \bibnamefont{Gambetta}},
  \bibinfo{author}{\bibfnamefont{L.}~\bibnamefont{Frunzio}},
  \bibinfo{author}{\bibfnamefont{S.~M.} \bibnamefont{Girvin}},
  \bibinfo{author}{\bibfnamefont{M.~H.} \bibnamefont{Devoret}},
  \bibnamefont{and} \bibinfo{author}{\bibfnamefont{R.~J.}
  \bibnamefont{Schoelkopf}}, \bibinfo{journal}{Nature (London)}
  \textbf{\bibinfo{volume}{467}}, \bibinfo{pages}{574} (\bibinfo{year}{2010}).

\bibitem[{\citenamefont{Lupascu \emph{et~al.}}(2007)\citenamefont{Lupascu,
  Saito, Picot, de~Groot, Harmans, and Mooij}}]{LupascuQND07}
\bibinfo{author}{\bibfnamefont{A.}~\bibnamefont{Lupascu}},
  \bibinfo{author}{\bibfnamefont{S.}~\bibnamefont{Saito}},
  \bibinfo{author}{\bibfnamefont{T.}~\bibnamefont{Picot}},
  \bibinfo{author}{\bibfnamefont{P.~C.} \bibnamefont{de~Groot}},
  \bibinfo{author}{\bibfnamefont{C.~J. P.~M.} \bibnamefont{Harmans}},
  \bibnamefont{and} \bibinfo{author}{\bibfnamefont{J.~E.} \bibnamefont{Mooij}},
  \bibinfo{journal}{Nature Physics} \textbf{\bibinfo{volume}{3}},
  \bibinfo{pages}{119} (\bibinfo{year}{2007}).

\bibitem[{\citenamefont{Johnson \emph{et~al.}}(2010)\citenamefont{Johnson,
  Reed, Houck, Schuster, Bishop, Ginossar, Gambetta, DiCarlo, Frunzio, Girvin,
  and Schoelkopf}}]{Johnson_NaturePhys10}
\bibinfo{author}{\bibfnamefont{B.~R.} \bibnamefont{Johnson}},
  \bibinfo{author}{\bibfnamefont{M.~D.} \bibnamefont{Reed}},
  \bibinfo{author}{\bibfnamefont{A.~A.} \bibnamefont{Houck}},
  \bibinfo{author}{\bibfnamefont{D.~I.} \bibnamefont{Schuster}},
  \bibinfo{author}{\bibfnamefont{L.~S.} \bibnamefont{Bishop}},
  \bibinfo{author}{\bibfnamefont{E.}~\bibnamefont{Ginossar}},
  \bibinfo{author}{\bibfnamefont{J.~M.} \bibnamefont{Gambetta}},
  \bibinfo{author}{\bibfnamefont{L.}~\bibnamefont{DiCarlo}},
  \bibinfo{author}{\bibfnamefont{L.}~\bibnamefont{Frunzio}},
  \bibinfo{author}{\bibfnamefont{S.~M.} \bibnamefont{Girvin}},
  \bibnamefont{and} \bibinfo{author}{\bibfnamefont{R.~J.}
  \bibnamefont{Schoelkopf}}, \bibinfo{journal}{Nature Physics}
  \textbf{\bibinfo{volume}{6}}, \bibinfo{pages}{663} (\bibinfo{year}{2010}).

\bibitem[{\citenamefont{Koch \emph{et~al.}}(2007)\citenamefont{Koch, Yu,
  Gambetta, Houck, Schuster, Majer, Blais, Devoret, Girvin, and
  Schoelkopf}}]{Koch_TransmonPRA07}
\bibinfo{author}{\bibfnamefont{J.}~\bibnamefont{Koch}},
  \bibinfo{author}{\bibfnamefont{T.~M.} \bibnamefont{Yu}},
  \bibinfo{author}{\bibfnamefont{J.}~\bibnamefont{Gambetta}},
  \bibinfo{author}{\bibfnamefont{A.~A.} \bibnamefont{Houck}},
  \bibinfo{author}{\bibfnamefont{D.~I.} \bibnamefont{Schuster}},
  \bibinfo{author}{\bibfnamefont{J.}~\bibnamefont{Majer}},
  \bibinfo{author}{\bibfnamefont{A.}~\bibnamefont{Blais}},
  \bibinfo{author}{\bibfnamefont{M.~H.} \bibnamefont{Devoret}},
  \bibinfo{author}{\bibfnamefont{S.~M.} \bibnamefont{Girvin}},
  \bibnamefont{and} \bibinfo{author}{\bibfnamefont{R.~J.}
  \bibnamefont{Schoelkopf}}, \bibinfo{journal}{Phys. Rev. A}
  \textbf{\bibinfo{volume}{76}}, \bibinfo{pages}{042319}
  (\bibinfo{year}{2007}).

\bibitem[{\citenamefont{Schreier \emph{et~al.}}(2008)\citenamefont{Schreier,
  Houck, Koch, Schuster, Johnson, Chow, Gambetta, Majer, Frunzio, Devoret,
  Girvin, and Schoelkopf}}]{SchreierPRB08}
\bibinfo{author}{\bibfnamefont{J.~A.} \bibnamefont{Schreier}},
  \bibinfo{author}{\bibfnamefont{A.~A.} \bibnamefont{Houck}},
  \bibinfo{author}{\bibfnamefont{J.}~\bibnamefont{Koch}},
  \bibinfo{author}{\bibfnamefont{D.~I.} \bibnamefont{Schuster}},
  \bibinfo{author}{\bibfnamefont{B.~R.} \bibnamefont{Johnson}},
  \bibinfo{author}{\bibfnamefont{J.~M.} \bibnamefont{Chow}},
  \bibinfo{author}{\bibfnamefont{J.~M.} \bibnamefont{Gambetta}},
  \bibinfo{author}{\bibfnamefont{J.}~\bibnamefont{Majer}},
  \bibinfo{author}{\bibfnamefont{L.}~\bibnamefont{Frunzio}},
  \bibinfo{author}{\bibfnamefont{M.~H.} \bibnamefont{Devoret}},
  \bibinfo{author}{\bibfnamefont{S.~M.} \bibnamefont{Girvin}},
  \bibnamefont{and} \bibinfo{author}{\bibfnamefont{R.~J.}
  \bibnamefont{Schoelkopf}}, \bibinfo{journal}{Phys. Rev. B}
  \textbf{\bibinfo{volume}{77}}, \bibinfo{pages}{180502}
  (\bibinfo{year}{2008}).

\bibitem[{\citenamefont{Vijay \emph{et~al.}}(2011)\citenamefont{Vijay,
  Slichter, and Siddiqi}}]{Vijay_PRL11}
\bibinfo{author}{\bibfnamefont{R.}~\bibnamefont{Vijay}},
  \bibinfo{author}{\bibfnamefont{D.~H.} \bibnamefont{Slichter}},
  \bibnamefont{and} \bibinfo{author}{\bibfnamefont{I.}~\bibnamefont{Siddiqi}},
  \bibinfo{journal}{Phys. Rev. Lett.} \textbf{\bibinfo{volume}{106}},
  \bibinfo{pages}{110502} (\bibinfo{year}{2011}).

\bibitem[{\citenamefont{Houck \emph{et~al.}}(2008)\citenamefont{Houck,
  Schreier, Johnson, Chow, Koch, Gambetta, Schuster, Frunzio, Devoret, Girvin,
  and Schoelkopf}}]{Houck_PRL08}
\bibinfo{author}{\bibfnamefont{A.~A.} \bibnamefont{Houck}},
  \bibinfo{author}{\bibfnamefont{J.~A.} \bibnamefont{Schreier}},
  \bibinfo{author}{\bibfnamefont{B.~R.} \bibnamefont{Johnson}},
  \bibinfo{author}{\bibfnamefont{J.~M.} \bibnamefont{Chow}},
  \bibinfo{author}{\bibfnamefont{J.}~\bibnamefont{Koch}},
  \bibinfo{author}{\bibfnamefont{J.~M.} \bibnamefont{Gambetta}},
  \bibinfo{author}{\bibfnamefont{D.~I.} \bibnamefont{Schuster}},
  \bibinfo{author}{\bibfnamefont{L.}~\bibnamefont{Frunzio}},
  \bibinfo{author}{\bibfnamefont{M.~H.} \bibnamefont{Devoret}},
  \bibinfo{author}{\bibfnamefont{S.~M.} \bibnamefont{Girvin}},
  \bibnamefont{and} \bibinfo{author}{\bibfnamefont{R.~J.}
  \bibnamefont{Schoelkopf}}, \bibinfo{journal}{Phys. Rev. Lett.}
  \textbf{\bibinfo{volume}{101}}, \bibinfo{pages}{080502}
  (\bibinfo{year}{2008}).

\bibitem[{\citenamefont{Dolan}(1977)}]{Dolan77}
\bibinfo{author}{\bibfnamefont{G.~J.} \bibnamefont{Dolan}},
  \bibinfo{journal}{Appl. Phys. Lett.} \textbf{\bibinfo{volume}{31}},
  \bibinfo{pages}{337} (\bibinfo{year}{1977}).

\bibitem[{\citenamefont{Simmonds \emph{et~al.}}(2004)\citenamefont{Simmonds,
  Lang, Hite, Nam, Pappas, and Martinis}}]{SimmondsPRL04}
\bibinfo{author}{\bibfnamefont{R.~W.} \bibnamefont{Simmonds}},
  \bibinfo{author}{\bibfnamefont{K.~M.} \bibnamefont{Lang}},
  \bibinfo{author}{\bibfnamefont{D.~A.} \bibnamefont{Hite}},
  \bibinfo{author}{\bibfnamefont{S.}~\bibnamefont{Nam}},
  \bibinfo{author}{\bibfnamefont{D.~P.} \bibnamefont{Pappas}},
  \bibnamefont{and} \bibinfo{author}{\bibfnamefont{J.~M.}
  \bibnamefont{Martinis}}, \bibinfo{journal}{Phys. Rev. Lett.}
  \textbf{\bibinfo{volume}{93}}, \bibinfo{pages}{077003}
  (\bibinfo{year}{2004}).

\bibitem[{\citenamefont{Martinis \emph{et~al.}}(2005)\citenamefont{Martinis,
  Cooper, McDermott, Steffen, Ansmann, Osborn, Cicak, Oh, Pappas, Simmonds, and
  Yu}}]{MartinisPRL05}
\bibinfo{author}{\bibfnamefont{J.~M.} \bibnamefont{Martinis}},
  \bibinfo{author}{\bibfnamefont{K.~B.} \bibnamefont{Cooper}},
  \bibinfo{author}{\bibfnamefont{R.}~\bibnamefont{McDermott}},
  \bibinfo{author}{\bibfnamefont{M.}~\bibnamefont{Steffen}},
  \bibinfo{author}{\bibfnamefont{M.}~\bibnamefont{Ansmann}},
  \bibinfo{author}{\bibfnamefont{K.~D.} \bibnamefont{Osborn}},
  \bibinfo{author}{\bibfnamefont{K.}~\bibnamefont{Cicak}},
  \bibinfo{author}{\bibfnamefont{S.}~\bibnamefont{Oh}},
  \bibinfo{author}{\bibfnamefont{D.~P.} \bibnamefont{Pappas}},
  \bibinfo{author}{\bibfnamefont{R.~W.} \bibnamefont{Simmonds}},
  \bibnamefont{and} \bibinfo{author}{\bibfnamefont{C.~C.} \bibnamefont{Yu}},
  \bibinfo{journal}{Phys. Rev. Lett.} \textbf{\bibinfo{volume}{95}},
  \bibinfo{pages}{210503} (\bibinfo{year}{2005}).

\bibitem[{\citenamefont{Kline \emph{et~al.}}(2009)\citenamefont{Kline, Wang,
  Oh, Martinis, and Pappas}}]{KlineSST09}
\bibinfo{author}{\bibfnamefont{J.~S.} \bibnamefont{Kline}},
  \bibinfo{author}{\bibfnamefont{H.}~\bibnamefont{Wang}},
  \bibinfo{author}{\bibfnamefont{S.}~\bibnamefont{Oh}},
  \bibinfo{author}{\bibfnamefont{J.~M.} \bibnamefont{Martinis}},
  \bibnamefont{and} \bibinfo{author}{\bibfnamefont{D.~P.}
  \bibnamefont{Pappas}}, \bibinfo{journal}{Supercond. Sci. Technol.}
  \textbf{\bibinfo{volume}{22}}, \bibinfo{pages}{015004}
  (\bibinfo{year}{2009}).

\bibitem[{\citenamefont{Oh \emph{et~al.}}(2006{\natexlab{a}})\citenamefont{Oh,
  Cicak, Kline, Sillanp\"{a}\"{a}, Osborn, Whittaker, Simmonds, and
  Pappas}}]{Oh_PRB06}
\bibinfo{author}{\bibfnamefont{S.}~\bibnamefont{Oh}},
  \bibinfo{author}{\bibfnamefont{K.}~\bibnamefont{Cicak}},
  \bibinfo{author}{\bibfnamefont{J.~S.} \bibnamefont{Kline}},
  \bibinfo{author}{\bibfnamefont{M.~A.} \bibnamefont{Sillanp\"{a}\"{a}}},
  \bibinfo{author}{\bibfnamefont{K.~D.} \bibnamefont{Osborn}},
  \bibinfo{author}{\bibfnamefont{J.~D.} \bibnamefont{Whittaker}},
  \bibinfo{author}{\bibfnamefont{R.~W.} \bibnamefont{Simmonds}},
  \bibnamefont{and} \bibinfo{author}{\bibfnamefont{D.~P.}
  \bibnamefont{Pappas}}, \bibinfo{journal}{Phys. Rev. B}
  \textbf{\bibinfo{volume}{74}}, \bibinfo{pages}{100502}
  (\bibinfo{year}{2006}{\natexlab{a}}).

\bibitem[{\citenamefont{Weides \emph{et~al.}}(2011)\citenamefont{Weides,
  Bialczak, Lenander, Lucero, Mariantoni, Neeley, O'Connell, Sank, Wang,
  Wenner, Yamamoto, Yin, Cleland, and Martinis}}]{Weides_Trilayer}
\bibinfo{author}{\bibfnamefont{M.}~\bibnamefont{Weides}},
  \bibinfo{author}{\bibfnamefont{R.~C.} \bibnamefont{Bialczak}},
  \bibinfo{author}{\bibfnamefont{M.}~\bibnamefont{Lenander}},
  \bibinfo{author}{\bibfnamefont{E.}~\bibnamefont{Lucero}},
  \bibinfo{author}{\bibfnamefont{M.}~\bibnamefont{Mariantoni}},
  \bibinfo{author}{\bibfnamefont{M.}~\bibnamefont{Neeley}},
  \bibinfo{author}{\bibfnamefont{A.~D.} \bibnamefont{O'Connell}},
  \bibinfo{author}{\bibfnamefont{D.}~\bibnamefont{Sank}},
  \bibinfo{author}{\bibfnamefont{H.}~\bibnamefont{Wang}},
  \bibinfo{author}{\bibfnamefont{J.}~\bibnamefont{Wenner}},
  \bibinfo{author}{\bibfnamefont{T.}~\bibnamefont{Yamamoto}},
  \bibinfo{author}{\bibfnamefont{Y.}~\bibnamefont{Yin}},
  \bibinfo{author}{\bibfnamefont{A.~N.} \bibnamefont{Cleland}},
  \bibnamefont{and} \bibinfo{author}{\bibfnamefont{J.}~\bibnamefont{Martinis}},
  \bibinfo{journal}{Supercond. Sci. Technol.} \textbf{\bibinfo{volume}{24}},
  \bibinfo{pages}{055005} (\bibinfo{year}{2011}).

\bibitem[{\citenamefont{Oh \emph{et~al.}}(2006{\natexlab{b}})\citenamefont{Oh,
  Hite, Cicak, Osborn, Simmonds, McDermott, Cooper, Steffen, Martinis, and
  Pappas}}]{OhEpitaxialReSputtering}
\bibinfo{author}{\bibfnamefont{S.}~\bibnamefont{Oh}},
  \bibinfo{author}{\bibfnamefont{D.~A.} \bibnamefont{Hite}},
  \bibinfo{author}{\bibfnamefont{K.}~\bibnamefont{Cicak}},
  \bibinfo{author}{\bibfnamefont{K.~D.} \bibnamefont{Osborn}},
  \bibinfo{author}{\bibfnamefont{R.~W.} \bibnamefont{Simmonds}},
  \bibinfo{author}{\bibfnamefont{R.}~\bibnamefont{McDermott}},
  \bibinfo{author}{\bibfnamefont{K.~B.} \bibnamefont{Cooper}},
  \bibinfo{author}{\bibfnamefont{M.}~\bibnamefont{Steffen}},
  \bibinfo{author}{\bibfnamefont{J.~M.} \bibnamefont{Martinis}},
  \bibnamefont{and} \bibinfo{author}{\bibfnamefont{D.~P.}
  \bibnamefont{Pappas}}, \bibinfo{journal}{Thin Solid Films}
  \textbf{\bibinfo{volume}{496}}, \bibinfo{pages}{389}
  (\bibinfo{year}{2006}{\natexlab{b}}).

\bibitem[{\citenamefont{Kline \emph{et~al.}}(2011)\citenamefont{Kline, Vissers,
  da~Silva, Wisbey, Weides, Shalibo, Katz, Johnson, Ohki, and
  Pappas}}]{Kline_SUST11}
\bibinfo{author}{\bibfnamefont{J.~S.} \bibnamefont{Kline}},
  \bibinfo{author}{\bibfnamefont{M.~R.} \bibnamefont{Vissers}},
  \bibinfo{author}{\bibfnamefont{F.~C.~S.} \bibnamefont{da~Silva}},
  \bibinfo{author}{\bibfnamefont{D.~S.} \bibnamefont{Wisbey}},
  \bibinfo{author}{\bibfnamefont{M.}~\bibnamefont{Weides}},
  \bibinfo{author}{\bibfnamefont{Y.}~\bibnamefont{Shalibo}},
  \bibinfo{author}{\bibfnamefont{N.}~\bibnamefont{Katz}},
  \bibinfo{author}{\bibfnamefont{B.~R.} \bibnamefont{Johnson}},
  \bibinfo{author}{\bibfnamefont{T.~A.} \bibnamefont{Ohki}}, \bibnamefont{and}
  \bibinfo{author}{\bibfnamefont{D.~P.} \bibnamefont{Pappas}},
  \bibinfo{journal}{arXiv:1108.1830v1}  (\bibinfo{year}{2011}).

\bibitem[{\citenamefont{C\'{o}rcoles
  \emph{et~al.}}(2011)\citenamefont{C\'{o}rcoles, Chow, Gambetta, Rigetti,
  Rozen, Keefe, Rothwell, Ketchen, and Steffen}}]{Corcoles_APL_2011}
\bibinfo{author}{\bibfnamefont{A.~D.} \bibnamefont{C\'{o}rcoles}},
  \bibinfo{author}{\bibfnamefont{J.~M.} \bibnamefont{Chow}},
  \bibinfo{author}{\bibfnamefont{J.~M.} \bibnamefont{Gambetta}},
  \bibinfo{author}{\bibfnamefont{C.}~\bibnamefont{Rigetti}},
  \bibinfo{author}{\bibfnamefont{J.~R.} \bibnamefont{Rozen}},
  \bibinfo{author}{\bibfnamefont{G.~A.} \bibnamefont{Keefe}},
  \bibinfo{author}{\bibfnamefont{M.~B.} \bibnamefont{Rothwell}},
  \bibinfo{author}{\bibfnamefont{M.~B.} \bibnamefont{Ketchen}},
  \bibnamefont{and} \bibinfo{author}{\bibfnamefont{M.}~\bibnamefont{Steffen}},
  \bibinfo{journal}{Appl. Phys. Lett.} \textbf{\bibinfo{volume}{99}},
  \bibinfo{pages}{181906} (\bibinfo{year}{2011}).

\bibitem[{\citenamefont{Allman \emph{et~al.}}(2010)\citenamefont{Allman,
  Altomare, Whittaker, Cicak, Li, Sirois, Strong, Teufel, and
  Simmonds}}]{AllmanArXiv}
\bibinfo{author}{\bibfnamefont{M.~S.} \bibnamefont{Allman}},
  \bibinfo{author}{\bibfnamefont{F.}~\bibnamefont{Altomare}},
  \bibinfo{author}{\bibfnamefont{J.~D.} \bibnamefont{Whittaker}},
  \bibinfo{author}{\bibfnamefont{K.}~\bibnamefont{Cicak}},
  \bibinfo{author}{\bibfnamefont{D.}~\bibnamefont{Li}},
  \bibinfo{author}{\bibfnamefont{A.}~\bibnamefont{Sirois}},
  \bibinfo{author}{\bibfnamefont{J.}~\bibnamefont{Strong}},
  \bibinfo{author}{\bibfnamefont{J.~D.} \bibnamefont{Teufel}},
  \bibnamefont{and} \bibinfo{author}{\bibfnamefont{R.~W.}
  \bibnamefont{Simmonds}}, \bibinfo{journal}{arXiv:1004.2738}
  (\bibinfo{year}{2010}).

\end{thebibliography}
\end{document}